\documentclass[aps,prl,reprint,superscriptaddress,showpacs]{revtex4-1}

\usepackage{graphicx}
\usepackage{amssymb,amsmath}
\usepackage{bm}
\usepackage{dcolumn}

\begin{document}

\title{Ultracold collision in the presence of synthetic spin-orbit coupling}

\author{Hao Duan}
\affiliation{State Key Laboratory of Low-Dimensional Quantum Physics, 
	Department of Physics, Tsinghua University, 
	Beijing 100084, 
	China}
\author{Li You}
\email[]{lyou@mail.tsinghua.edu.cn}
\affiliation{State Key Laboratory of Low-Dimensional Quantum Physics, 
	Department of Physics, Tsinghua University, 
	Beijing 100084, 
	China}
\author{Bo Gao}
\email[]{bo.gao@utoledo.edu}
\affiliation{Department of Physics and Astronomy,
	Mailstop 111,
	University of Toledo,
	Toledo, Ohio 43606,
	USA}

\date{January 24, 2013}

\begin{abstract}

We present an analytic description of ultracold collision between 
two spin-$\frac{1}{2}$ fermions with isotropic spin-orbit coupling of the Rashba type.
We show that regardless of how weak the spin-orbit coupling may be, the
ultracold collision at sufficiently low energies is significantly modified,
including the ubiquitous Wigner threshold behavior. We further show
that the particles are preferably scattered into the lower-energy helicity
state due to the break of parity conservation, thus establishing interaction
with spin-orbit coupling as one mechanism for the spontaneous emergence of handedness.

\end{abstract}


\maketitle


Systems of cold atoms have become fertile laboratories for many-body
and few-body physics largely because of the ability to tune and manipulate atomic interactions.
The magnetic Feshbach resonance \cite{chi10}, for instance, has allowed precise tuning of
scattering length to virtually arbitrary value, facilitating studies
of strongly coupled many-body systems \cite{Bloch2008,Giorgini2008}
and also few-body systems in the
universal regime \cite{kra06,bra06,gre10}.

A new class of manipulation of cold atoms has arisen recently under the
general envelope of synthetic gauge fields,
generated mainly through coherent laser-atom interactions \cite{dal11}.
Among various types, the synthetic spin-orbit coupling (SOC) 
\cite{Lin2011,Wang2012,Cheuk2012,Zhang2012,Qu2013} is of special interest
as it simulates a type of coupling that is regarded as important in 
fractional quantum Hall effect
and topological insulators \cite{Hasan2010,Qi2011}.
Despite a large body of recent works on SOC systems 
\cite{Ho2011,Vyasanakere2011a,Xu2011,Vyasanakere2011b,Gong2011,Yu2011,Hu2011,
Xu2012a,Anderson2012,Xu2012b,Vyasanakere2012}, 
many fundamental questions remain, as elementary as effects of
SOC on the two-body scattering \cite{Cui2012,Zhang2012a,Zhang2012b}. 
A recent experiment by Williams \textit{et al.} \cite{Williams2012}
provides an early indication that such effects can be substantial,
and essential for understanding interacting many-body, and few-body 
systems with SOC.

In this work, we present a general theoretical treatment of the scattering of
two spin-$\frac{1}{2}$ fermions with isotropic SOC of the Rashba type.
We pick spin-$\frac{1}{2}$ fermions for its direct relevance to electrons in condensed matter,
and for the fact that it can be simulated by $^6$Li in its ground hyperfine
state \cite{Cheuk2012}, or in the ultracold regime by pseudo 
spin-$\frac{1}{2}$ fermions \cite{Wang2012}. 
We choose isotropic coupling to isolate effects of SOC and effects of anisotropy. 
We show that in the presence of SOC, a non-Abelian gauge field that persists to
infinite inter-particle separation, the scattering formulation has to be changed substantially,
including the very definitions of fundamental quantities such as 
the scattering matrices and the flux. 
The formalism is solved analytically, in terms of scattering in the absence 
of SOC, by taking advantage of a length scale separation \cite{gao05a,che07}.
The results, when compared with ones without SOC,
show a substantially altered threshold behavior, different from the familiar
Wigner behavior \cite{wig48}, and a preferential
scattering into the lower-energy helicity state as a consequence of parity 
non-conservation. This preference implies that handedness can spontaneously
emerge as a result of scattering with SOC.


We consider two identical particles with $F_1=F_2=1/2$. 
We use symbols $F_1$ and $F_2$ to distinguish them, for atoms, 
from electronic spin which has a well-defined separate meaning.
In the absence of SOC, the interaction between two such particles can 
very generally be described by the Hamiltonian
\begin{equation}
H = h_1+h_2+\widehat{V} \;.
\end{equation}
Here $h_i=\bm{p}_i^2/2m$ is the single particle
Hamiltonian in the absence of SOC, and $\widehat{V}$ is an interaction
operator describing two effective central potentials,
$V^{(F=0)}(r)$ for the ``singlet'' states and $V^{(F=1)}(r)$ for the 
``triplet'' states. Without SOC, the total ``spin'', $\bm{F}=\bm{F}_1+\bm{F}_2$, 
and the orbital angular momentum
$\bm{l}$ are independently conserved. 
The scattering is fully characterized by two sets of effective single-channel
$K$ matrices, $\tan\delta_l^{F=0}$ for the singlet states
and $\tan\delta_l^{F=1}$ for the triplet states. 

\begin{figure}
\includegraphics[width=\columnwidth]{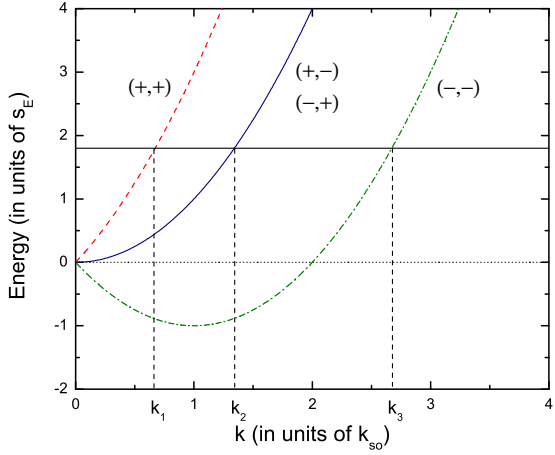}
\caption{(color online) The three branches of dispersion relation for
two particles with SOC in the center-of-mass frame.
For each energy $\epsilon=\hbar^2 k^2/2\mu>0$, there are three 
corresponding $k$'s, which, in the order of increasing magnitude,
are given by $k_1=\sqrt{k_\mathrm{so}^2+k^2}-k_\mathrm{so}$ for the
$(+,+)$ channel, $k_2=k$ for the $(+,-)$ and $(-,+)$ channels, and
$k_3=\sqrt{k_\mathrm{so}^2+k^2}+k_\mathrm{so}$ for the $(-,-)$ channel.
Here $s_E \equiv \hbar^2k_\mathrm{so}^2/2\mu$ is the energy scale for the SOC.
\label{fig:disp}}
\end{figure}

The isotropic SOC of the Rashba type changes the single particle
Hamiltonian from $h_i=\bm{p}_i^2/2m$ to 
\begin{equation}
h_i=\frac{\bm{p}_i^{2}}{2m}
+\frac{\hbar C_\mathrm{so}}{m}\,\bm{\sigma}_i\cdot\bm{p}_i,
\label{single-H}
\end{equation}
where $\bm{\sigma}$ denotes the Pauli spin matrix,
$C_\mathrm{so}$ is a constant characterizing the strength of SOC.
It has a dimension of a $k$-vector (inverse length), with its
magnitude to be denoted by $k_\mathrm{so}\equiv |C_\mathrm{so}|$.
This single particle Hamiltonian is diagonalized by states
$|\pm,\bm{n}_\mathrm{so}\rangle_i|\bm{K}_i\rangle$, where
$|\bm{K}_i\rangle$ describes the translational motion,
and is an eigenstate of $\bm{p}_i$ with an eigenvalue of $\hbar\bm{K}_i$.
$|\pm,\bm{n}_\mathrm{so}\rangle_i$ is a short-hand notation for
$|F_i=1/2,M_i=\pm 1/2,\bm{n}_\mathrm{so}\rangle_i$ with
$\bm{n}_\mathrm{so}$ defining the direction of quantization.
For isotropic SOC of the Rashba type, $\bm{n}_\mathrm{so}=\bm{K}_i/K_i$ 
for $C_\mathrm{so}>0$, and $\bm{n}_\mathrm{so}=-\bm{K}_i/K_i$ 
for $C_\mathrm{so}<0$. The ``$\pm$'' states have different energies
as given by two distinctive dispersion relations, 
$E_i = \hbar^2K_i^2/2m\pm \hbar^2k_\mathrm{so}K_i/m$. We will generally refer to 
the ``$\pm$'' states as the helicity states, or the ``$+$'' state as the ``higher-energy''
helicity state, the ``$-$'' state as the ``lower-energy''
helicity state, when more precision is required.

For two interacting particles with SOC, the conservation of the
total momentum, $\bm{P}=\bm{p}_1+\bm{p}_2$, allows the investigation
of scattering and interaction in the center-of-mass frame, $\bm{P}=0$,
in which the relative motion is described by the Hamiltonian,
\begin{equation}
H_\mathrm{rel} = \frac{1}{2\mu}\bm{p}^2+\widehat{V}
	+\frac{\hbar C_\mathrm{so}}{m}\,(\bm{\sigma}_1-\bm{\sigma}_2)\cdot\bm{p} \;.
\end{equation}
Here $\mu=m/2$ is the reduced mass, and $\bm{p}=(\bm{p}_1-\bm{p}_2)/2$ is the
(canonical) momentum corresponding to the relative motion.
In the center-of-mass frame, SOC changes the
the single dispersion relation, $\epsilon=\hbar^2k^2/2\mu$, 
applicable to all spin states without SOC,
into three branches: $\epsilon=(\hbar^2/2\mu)(k^2+2k_\mathrm{so}k)$ 
for the $|+,\bm{n}_\mathrm{so}\rangle_1|+,-\bm{n}_\mathrm{so}\rangle_2$ 
two-particle spin state, where $\bm{n}_\mathrm{so}=\bm{k}/k$ 
for $C_\mathrm{so}>0$, and $\bm{n}_\mathrm{so}=-\bm{k}/k$ 
for $C_\mathrm{so}<0$,
$\epsilon=\hbar^2k^2/2\mu$ for the 
$|+,\bm{n}_\mathrm{so}\rangle_1|-,-\bm{n}_\mathrm{so}\rangle_2$ and
$|-,\bm{n}_\mathrm{so}\rangle_1|+,-\bm{n}_\mathrm{so}\rangle_2$ spin states,
and $\epsilon=(\hbar^2/2\mu)(k^2-2k_\mathrm{so}k)$ for the 
$|-,\bm{n}_\mathrm{so}\rangle_1|-,-\bm{n}_\mathrm{so}\rangle_2$ spin state.
This change of dispersion is one of the key characteristics of
interaction with SOC, and is illustrated in Fig.~\ref{fig:disp}.
The four spin states $|\pm,\bm{n}_\mathrm{so}\rangle_1|\pm,-\bm{n}_\mathrm{so}\rangle_2$ 
constitute what we call the two-particle helicity basis, 
and will be abbreviated as
$(+,+)$ ,$(+,-)$, $(-,+)$, and $(-,-)$, respectively.
They define the asymptotic channels for interaction with SOC.
The different dispersion relations for the $(+,+)$ and $(-,-)$
asymptotic states, which are related to each other by a parity
operation (and exchange of particles), are direct consequences
of the parity non-conservative nature of the SOC. The time-reversal
symmetry is, however, still maintained.

With SOC, even the isotropic SOC under consideration here, 
the $\bm{F}$ and $\bm{l}$ are generally no longer independently
conserved. For isotropic SOC, 
the total angular momentum $\bm{F}_t=\bm{F}+\bm{l}$ is conserved.
The wave function for each total angular momentum, $F_tM_t$, can be
expanded as
\[
\psi^{F_tM_t}_j = \sum_{\alpha}\Phi^{F_tM_t}_{\alpha}G^{F_t}_{\alpha j}(r)/r \;.
\]
Here $G^{F_t}_{\alpha}/r$ describes the relative radial motion, and $j$ is an index for different
linearly independent solutions. The $\Phi^{F_tM_t}_{\alpha}$ are channel functions,
indexed by $\alpha$, describing all degrees of
freedom other than the relative radial motion. They are conveniently
chosen here to be the $\{F,l\}$ basis, in which the interaction in the absence of SOC is
diagonal. The summation over $\alpha$, namely the $F$ and $l$ combinations, 
is restricted both by the angular momentum conservation and by 
$F+l=\mathrm{even}$ as imposed by
the symmetry under the exchange of particles \cite{gao96}.
This leads to the following general channel structure for interaction
with isotropic SOC.
All $F_t=\mathrm{odd}$ states are described by single-channel problems
with $V^{(F=1)}(r)$, corresponding to $\{F=1,l=F_t\}$.
All $F_t=\mathrm{even}$ states, other than $F_t=0$, are described 
by three-channel problems, corresponding to $\{F=0,l=F_t\}$,
$\{F=1,l=F_t-1\}$, and $\{F=1,l=F_t+1\}$.
The $F_t=0$ states are described by a two-channel problem
with $\{F=0,l=0\}$ and $\{F=1,l=1\}$.
The radial functions $G^{F_t}_{\alpha}(r)$ for $F_t=\mathrm{even}$
satisfy coupled-channel equations, which are given explicitly, 
for $F_t=0$, by
\begin{widetext}
\begin{equation}
\left(
\begin{array}{cc}
-\frac{\hbar^{2}}{2\mu}\frac{d^{2}}{dr^{2}}+V^{(0)}(r)-\epsilon & 
	i2\frac{\hbar^2C_\mathrm{so}}{m}\left(\frac{d}{dr}+\frac{1}{r}\right) \\
i2\frac{\hbar^2C_\mathrm{so}}{m}\left(\frac{d}{dr}-\frac{1}{r}\right) &
	-\frac{\hbar^{2}}{2\mu}\frac{d^{2}}{dr^{2}}+\frac{2\hbar^{2}}{2\mu r^{2}}
	+V^{(1)}(r)-\epsilon
\end{array}
\right)
\left(
\begin{array}{c}
G^{F_t=0}_{F=0l=0}\\
G^{F_t=0}_{F=1l=1}
\end{array}
\right)=0 \;.
\label{eq:cc0}
\end{equation}
\end{widetext}
The persistence of the SOC, corresponding to the off-diagonal terms in
Eq.~(\ref{eq:cc0}), to infinite separation requires re-definitions of scattering matrices.
In particular, the $K$ matrix is now defined by
\begin{equation}
G^{F_t}/r \stackrel{r\rightarrow\infty}{\sim} 
	\mathcal{J}^{F_t} - \mathcal{Y}^{F_t}K^{F_t} \;,
\label{eq:Kdef}
\end{equation}
where
\begin{equation}
\mathcal{J}^{F_t=0} = 
\left(
\begin{array}{ccc}
\frac{1}{\sqrt{2}}k_{1}j_{0}(k_{1}r) & -\frac{1}{\sqrt{2}}k_{3}j_{0}(k_{3}r)\\
-i\frac{1}{\sqrt{2}}k_{1}j_{1}(k_{1}r) & -i\frac{1}{\sqrt{2}}k_{3}j_{1}(k_{3}r)
\end{array}
\right) \;,
\label{eq:J0}
\end{equation}
and
\begin{equation}
\mathcal{Y}^{F_t=0} =
\left(
\begin{array}{ccc}
\frac{1}{\sqrt{2}}k_{1}y_{0}(k_{1}r) & -\frac{1}{\sqrt{2}}k_{3}y_{0}(k_{3}r)\\
-i\frac{1}{\sqrt{2}}k_{1}y_{1}(k_{1}r) & -i\frac{1}{\sqrt{2}}k_{3}y_{1}(k_{3}r)
\end{array}
\right) \;.
\label{eq:Y0}
\end{equation}
Here $k_1=\sqrt{k_\mathrm{so}^2+k^2}-k_\mathrm{so}$ and
$k_3=\sqrt{k_\mathrm{so}^2+k^2}+k_\mathrm{so}$, as illustrated
in Fig.~\ref{fig:disp}, and $j_l(x)$ and $y_l(x)$ are the spherical
Bessel functions \cite{olv10}.
The $\mathcal{J}^{F_t=0}$ and $\mathcal{Y}^{F_t=0}$
are the exact regular and irregular analytic solutions of 
Eq.~(\ref{eq:cc0}) in the absence of interaction, namely for 
$V^{(0)}=V^{(1)}\equiv 0$. The two columns of the matrices correspond to
solutions for the $(+,+)$ and the $(-,-)$ channels, respectively.
Other scattering matrices such as the $S$ matrix can be 
defined in a similar manner with their usual relationships maintained.
For example, the $S$ matrix is related to the $K$ matrix by
$S^{F_t} = (I+iK^{F_t})(I-iK^{F_t})^{-1}$, where $I$ represents the unit matrix.
We note that in standard multichannel scattering theory without SOC
(see, e.g., Ref.~\cite{gao96}), $\mathcal{J}^{F_t}$ and $\mathcal{Y}^{F_t}$
would have been diagonal.

From linear superposition of solutions that define the scattering
matrices, a wave function satisfying scattering boundary condition
can be constructed, from which all physical observables as related
to scattering can be extracted \cite{gao96}. One obtains, for instance,
\begin{equation}
\sigma[(+,+)\!\to\!(+,+)] = \frac{2\pi}{k_1^2}
	\sum_{F_t=\mathrm{even}}(2F_t+1)\left|S^{F_t}_{11}-1\right|^2 \;,
\end{equation}
and similarly for other cross sections. We emphasize that
in deriving proper cross section formulas, it is crucial to recognize that
in the presence of a gauge field, the flux or the current
density has to be defined with the velocity operator (or the corresponding
\textit{kinetic} momentum)
$\bm{v} = \dot{\bm{r}} = [\bm{r},H_\mathrm{rel}]/i\hbar = 
\bm{p}/\mu+\hbar C_\mathrm{so}(\bm{\sigma}_1-\bm{\sigma}_2)/m$.
The $(+,+)$ and the $(-,-)$ states, despite having different 
\textit{canonical} momenta,
$\hbar k_1$ and $\hbar k_3$, respectively, have the same velocity of 
$\hbar\sqrt{k_\mathrm{so}^2+k^2}/\mu$. 
The $(+,-)$ and the $(-,+)$ states have a different, the standard velocity of
$\hbar k/\mu$. This difference in the definition of flux is
another general key ingredient for a proper description of 
scattering in a gauge field.


The above formulation for two spin-$\frac{1}{2}$ particles with SOC is very
general, applicable for arbitrary energy and SOC coupling strength. In reality, both experimental 
realizations of SOC \cite{Lin2011,Wang2012,Cheuk2012} and the very validity of the Hamiltonian
used to describe it, imply that we are most interested in a regime of
SOC being weak, in the following sense.
Let $r_0$ be the range of interaction without SOC, the energy scale associated
with SOC, $s_E \equiv \hbar^2k_\mathrm{so}^2/2\mu$, is generally much smaller that the energy
scale associated with the shorter-range interactions, $(\hbar^2/2\mu)(1/r_0^2)$, which
for atoms would be the van der Waals energy scale \cite{gao09a}.
This criterion, which is equivalent to a length scale separation,
$1/k_\mathrm{so}\gg r_0$, basically ensures that
the SOC and other interactions are important in different regions and
are not important simultaneously \cite{gao05a,che07}.
Under such a condition, scattering in the
presence of SOC can be solved in terms of scattering in the absence
of SOC.
Specifically, the $K$ matrix, as defined by Eq.~(\ref{eq:Kdef}),
can be obtained by matching Eq.~(\ref{eq:Kdef}), in a region of $r_0\ll r\ll 1/k_\mathrm{so}$,
to inner solutions for which the SOC is negligible.
This is conceptually similar to the multiscale quantum-defect treatment
of two atoms in a trap \cite{che07}.

For energies much greater than $s_E$, we obtain
the $K$ matrix to be given by the $K$ matrix in the $\{F,l\}$ basis through
a frame transformation. For $F_t=0$, e.g., we obtain
\begin{equation}
K^{F_t=0} = U^{F_t=0\dagger}
	\left(
	\begin{array}{cc}
	\tan\delta^{F=0}_{l=0} & 0\\
	0 & \tan\delta^{F=1}_{l=1}
	\end{array}
	\right)
	U^{F_t=0} \;,
\label{eq:K0noso}
\end{equation}
where $U^{F_t=0}$ is a global unitary matrix
\begin{equation}
U^{F_t=0} = \frac{1}{\sqrt{2}}
	\left(
	\begin{array}{cc}
	1 & -1\\
	-i & -i
	\end{array}
	\right) \;.
\end{equation}
This result, together with similar results for other total angular momenta,
has a very simple physical interpretation. It states that for 
energies much greater than the SOC energy scale, SOC has no
effect on the scattering dynamics, except to facilitate the preparation
and detection of particles in the helicity basis. 
In the absence of SOC, the same $K$ matrix describes scattering in
the helicity basis, and is applicable for all (positive) energies.

For energies comparable or smaller than $s_E$,
the length scale separation ensures that we are well into the region dominated 
by the $s$ wave scattering, which is well characterized,
for the vast majority of systems, by the
universal behaviors of $\tan\delta^{F=0}_{l=0}\approx-a_{F=0}k$
and $\tan\delta^{F=1}_{l=1}\approx 0$.
In this case, we obtain
\begin{equation}
K^{F_t=0} = -\frac{a_{F=0}}{k_3+k_1}\left(
\begin{array}{cc}
k_1^2 & -k_3k_1 \\
-k_3k_1 & k_3^2
\end{array}
\right) \;.
\label{eq:K0so}
\end{equation}
This result also represents the analytic solution of Eq.~(\ref{eq:cc0})
for the pseudopotential model \cite{Huang1957} of 
$V^{(0)}=\frac{4\pi\hbar^2a_{F=0}}{m}\delta(\bm{r})\frac{\partial}{\partial r}(r\cdot)$ 
and $V^{(1)}\equiv 0$, which is equivalent to imposing the boundary conditions
of $G^{F_t=0}_{F=0l=0}/r\stackrel{r\rightarrow 0}{\sim} A(1-a_{F=0}/r)$
and $G^{F_t=0}_{F=1l=1}/r\stackrel{r\rightarrow 0}{\sim} 0$.
The multiscale QDT approach contains the pseudopotential
results \cite{che07}. It is more general and leaves room for future generalizations,
including both the cases of non-universal behavior around $a_{F=0}=0$ \cite{gao09a}
and the case of much stronger SOC, the treatment of the latter would be
similar to the treatment of hyperfine effects in atomic scattering \cite{gao05a}.

The $K$ matrix of Eq.~(\ref{eq:K0so}) gives the following set of cross sections
for ultracold collision with SOC
\begin{subequations}
\begin{multline}
\sigma[(+,+)\!\to\!(+,+)] = \sigma[(-,-)\!\to\!(+,+)] \\
	= 8\pi a_{F=0}^{2}\frac{k_1^2}{(k_3+k_1)^2+a_{F=0}^{2}(k_3^2+k_1^2)^2} \;,
\end{multline}
\begin{multline}
\sigma[(-,-)\!\to\!(-,-)] = \sigma[(+,+)\!\to\!(-,-)] \\
	= 8\pi a_{F=0}^{2}\frac{k_3^2}{(k_3+k_1)^2+a_{F=0}^{2}(k_3^2+k_1^2)^2} \;.
\end{multline}
\label{eq:xsso}
\end{subequations}
In comparison, the cross sections in the absence of SOC, determined
by the $K$ matrix of Eq.~(\ref{eq:K0noso}) in the helicity basis, 
are given in the $s$ wave region by 
\begin{multline}
\sigma[(+,+)\!\to\!(+,+)] = \sigma[(+,+)\!\to\!(-,-)] \\
= \sigma[(-,-)\!\to\!(+,+)] = \sigma[(-,-)\!\to\!(-,-)] \\
= \frac{2\pi a_{F=0}^2}{1+a_{F=0}^2k^2} \;,
\label{eq:xsnoso}
\end{multline}
which all follow the Wigner threshold behavior \cite{wig48} of $\sigma\sim\mathrm{const.}$

\begin{figure}
\includegraphics[width=\columnwidth]{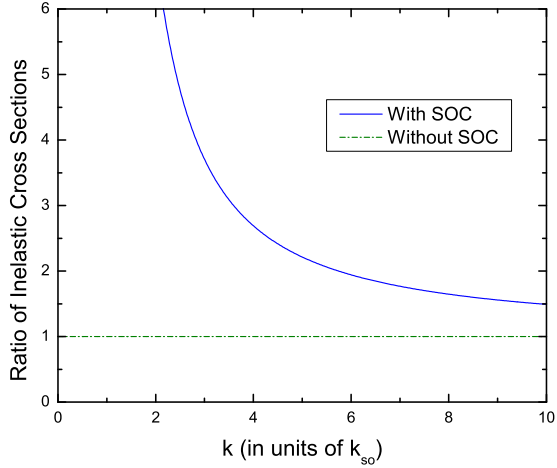}
\caption{(color online) The universal ratios of inelastic scattering cross sections,
$\sigma[(+,+)\!\to\!(-,-)]/\sigma[(-,-)\!\to\!(+,+)]$, 
with (solid line) and without SOC (dash-dot line), as a function of
$k$. The result with SOC is guaranteed
by the time-reversal symmetry to be valid at all energies.
The result without SOC is guaranteed by the combination of time-reversal 
and parity conservations at all energies.
The difference is due to the break of parity conservation by SOC.
\label{fig:xsRatio}}
\end{figure}

Equations~(\ref{eq:xsso})-(\ref{eq:xsnoso}) are the main results of this
work. They represent the universal behaviors followed by the vast majority
of spin-$\frac{1}{2}$ fermionic systems in the ultracold regime. The strength of SOC
only affects length and energy scaling, and with proper scaling, different
systems differ from each other only in a single dimensionless parameter
of $\eta_\mathrm{so}\equiv k_\mathrm{so}a_{F=0}$, with  $\eta_\mathrm{so}=\infty$
corresponding to the unitarity limit.

We focus here on two aspects of physics contained in these results. 
(a) The SOC has substantially
modified the threshold behavior, from the Wigner threshold law of
$\sigma\sim\mathrm{const.}$ for all cross sections, to
\begin{subequations}
\begin{multline}
\sigma[(+,+)\!\to\!(+,+)] = \sigma[(-,-)\!\to\!(+,+)] \\
	\sim \frac{\pi a_{F=0}^{2}}{2k_\mathrm{so}^{4}}k^{4} \;,
\end{multline}
\begin{multline}
\sigma[(-,-)\!\to\!(-,-)] = \sigma[(+,+)\!\to\!(-,-)] \\
	\sim 8\pi a_{F=0}^{2} \;,
\end{multline}
\end{subequations}
implying that the $(+,+)$ interaction is dominated by inelastic collision
into the $(-,-)$ channel, while $(-,-)$ interaction is dominated by elastic collision.
(b) Particles are preferably scattered into the lower-energy helicity state,
the ``$-$'' state, as reflected by 
$\sigma[(+,+)\!\to\!(-,-)]$ being always greater than
$\sigma[(-,-)\!\to\!(+,+)]$. More specifically
\begin{equation}
\frac{\sigma[(+,+)\!\to\!(-,-)]}{\sigma[(-,-)\!\to\!(+,+)]}
	=\frac{k_3^2}{k_1^2} = 
	\left(\frac{\sqrt{1+(k/k_\mathrm{so})^2}+1}
	{\sqrt{1+(k/k_\mathrm{so})^2}-1}\right)^2>1 \;,
\label{eq:xsRatio}
\end{equation}
and diverges as $1/k^4$ around the threshold. Equation~(\ref{eq:xsRatio}) for the ratio
of inelastic cross sections is applicable not only in the ultracold region, 
but at arbitrary energy as a result of time-reversal symmetry \cite{lan77}.
Its implication is best understood by noting that in the absence
of SOC, the two inelastic cross sections are strictly equal at all energies
as guaranteed by the combination of time-reversal and parity conservations.
The two universal ratios are compared in Fig.~\ref{fig:xsRatio}.
In an ultracold sample with SOC, the $(+,+)$ state has a finite cross section
to be converted into $(-,-)$, the $(+,-)$ and $(-,+)$ interactions are negligible,
and the $(-,-)$ state interacts mostly elastically, namely remains
in $(-,-)$. Independent of the initial statistical distribution, such a system
will evolve into a steady state made of mostly particles in 
the lower-energy helicity ``$-$'' state. In other words, a system with a preferred 
chirality would develop spontaneously through interaction.

In conclusion, we have developed a general formalism for the scattering 
of two spin-$\frac{1}{2}$ particles in the presence of isotropic SOC.
We believe it to be the first rigorous formulation for scattering
in a non-Abelian gauge field. 
We have derived the universal analytic results in the ultracold regime 
and discussed their implications.
Many of the concepts introduced are generally applicable, and provide
important guidance for investigations of other spin systems and 
anisotropic SOC. The theory is part of an essential foundation for understanding 
interacting many-body and few-body systems with SOC.


\begin{acknowledgments}
This work was supported by NSFC (No.~91121005 and No.~11004116),
MOST 2013CB922000 of the National Key Basic Research Program of China,
and the research program 2010THZO of the Tsinghua University.
\end{acknowledgments}

\bibliography{gauge,fewbody,bgao}

\end{document}